\documentclass{iaus}
\usepackage{graphicx}

\title[SMC: SFH and CEH] %% give here short title %%
{The Star Formation History in a SMC field: IAC-star/IAC-pop at work}

\author[No\"el et al.]   %% give here short author list %%
{Noelia E. D. No\"el$^1$,  Antonio Aparicio$^1$, Carme Gallart$^1$, 
Sebasti\'an L. Hidalgo$^2$,
 Edgardo Costa$^3$ \and Ren\'e A. M\'endez$^3$}

\affiliation{$^1$Instituto de Astrof\'\i sica de Canarias. 38200 La
Laguna. Tenerife, Canary Islands. Spain \break email: 
noelia@iac.es\\[\affilskip]$^2$University of Minnesota, 
Department of Astronomy, 
116 Church St. S.E., 
Minneapolis, MN 55455\\[\affilskip]
$^3$Departamento de Astronom\'\i a, Universidad de Chile, Casilla 36-D,
 Santiago, Chile}

\pubyear{2007}
\volume{241}  %% insert here IAU Symposium No.
\pagerange{119--126}
\date{?? and in revised form ??}
\jname{Stellar Populations as Building Blocks of Galaxies}
\editors{A. Vazdekis \& R. Peletier, eds.}
\begin{document}

\maketitle

\begin{abstract}

We present a progress report of a project to study the quantitative 
 star formation history (SFH)
in different parts of the Small Magellanic Cloud (SMC). We use the information 
in  [(B-R), R] color-magnitude diagrams (CMDs), which reach down to the oldest main-sequence
turnoffs and allow us to retrieve the SFH in detail.
We show the first results of the SFH in a SMC field located in the Southern direction (at $\thicksim$1 kpc from the SMC center).
This field is particularly interesting because in spite of being located in a place 
in which the HI column density is very low, it still presents a recent enhancement of star formation.  

\keywords{local group galaxies: individual (SMC) --- galaxies: star formation
history}
\end{abstract}

\firstsection % if your document starts with a section,
\section{Method}
Our study of the SFH is based on the method by Aparicio \& Hidalgo (2007, submitted).
 In short, once we obtained the CMD of our SMC field (No\"el et al. 2007), we prepared a synthetic CMD which 
 reproduce the magnitudes, ages, and metallicities of the stars of a given SFH. 
 This CMD, constructed using IAC-star (Aparicio \& Gallart 2004),
  contains the range of ages from 0 to 13 Gyr and  metallicities from 0.0001 to
   0.008,  
   an IMF from Kroupa et al. (2003) and 30\% of binary stars.
    In the analysis presented here, we used the BaSTI stellar evolution models
     (Pietrinferni et al. 2004).
The observational errors were simulated in the synthetic CMD following Gallart et al. (1999), to obtain the 
so-called model CMD. This model CMD was divided in simple populations of limited age and metallicity ranges. We choped the observed and
model CMDs into ``boxes'' using an  {\it {\`a} le carte} parameterization (see Aparicio 2007 et al., this conference) 
to see  how these simple and observed populations  
 are distributed. The distribution of stars in the defined boxes was calculated for any model SFH as a linear combination of
   the simple stellar populations. 
    Those areas of the CMD where the age-metallicity degeneracy is not significantly problematic and where the stars
    of different ages are more separated (MS), are better sampled 
    than those with more mixed populations.
Using the algorithm IAC-pop (see Aparicio et al. 2007, this conference), we compared  the model and synthetic CMDs and
 we found a suitable SFH.

\begin{figure}
\begin{center}

\includegraphics[height=4in,width=5in]{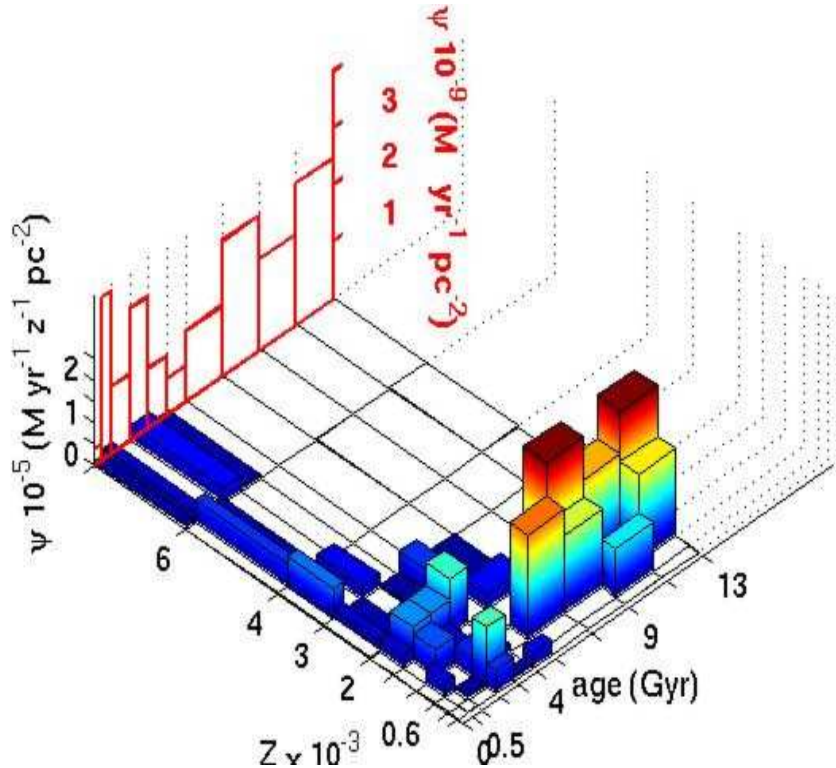}

\caption{SFH of SMC field smc0057 using the BaSTI library and {\it \`a le carte} 
 parameterization.
The x-axis represents the age in Gyr and y-axis the metallicity. The star formation rate, $\psi$,
 in units of solar masses per year, metallicity and area is given in the vertical axis.
 The SFH is given by the height of the bar emerging from the XY plane. The volume of
 this bar represents the number of stars formed in each interval of age and metallicity. 
 In the vertical
 x-axis the SFH is represented as a function of the age of the stars (integrated over the whole metallicity range). Error
 bars were omitted for clarity.}\label{fig:sfh}
\end{center}
\end{figure}

\section{Preliminary Results}
According to our solution (see Figure~\ref{fig:sfh}), this area of the SMC has been 
forming stars from ancient times until very recently, though some epochs were more active than others. 
A significant fraction of all the stars were formed before 7 Gyr ago. 
There is a recent enhancement of star formation, with some 5\% of stars younger than 0.5 Gyr.
It is important to remark that the HI structure of the SMC (Stanimirovi\'c et al. 1999)
 shows that the HI column density is very low in this part of the galaxy.
 The stars in this field cover the
range in metallicity from Z=0.0001-0.008. 
 The chemical enrichment is slow
 until $\thicksim$4 Gyr ago, and then it speeds up, in agreement with that found by Carrera (PhD Thesis) for this
SMC field using the CaII triplet (see No\"el et al. 2007, this conference).

%%%Given the location of this field, could be intuitive to think that the recent star formation is due to the high density concentration of 
%%%gas in this region of the galaxy. 
%%%However, the HI structure of the SMC (Stanimirovi\'c et al. 1999) shows that HI density column is very low here. Hence, the 
%%%recent star formation could be stired up by interactions between the LMC-SMC-Milky Way system.

%%%\subsection*{}   %%% Unnumbered second level section head (remove "%" symbol)

%%%\acknowledgements %%% Text of acknowledgements runs on after this command.

%%% THE BIBLIOGRAPHY


\begin{thebibliography}{}

\bibitem[Aparicio \& Gallart (2004)]{apagal04} 
      {Aparicio, A., \& Gallart, C.} 2004, 
       \textit{AJ} 128, 1465
     
\bibitem[Gallart et al. (1999)]{Gal99}
        {Gallart, C., Freedman, W. L., Aparicio, A.,
	 Bertelli, G., \& Chiosi, C.} 1999, 
	 \textit{AJ} 118, 2245

\bibitem[Kroupa et al. (2003)]{kroupa03} 
      {Kroupa, P, \& Weidner, C.} 2003, 
             \textit{ApJ} 598, 1076

\bibitem[Noel et al. (2007)]{noel07} 
      {No\"el, N. E. D., Gallart, G., Costa, E., \& M\'endez, R. A.} 2007, 
             \textit{AJ} in press

     
\bibitem[Teramo]{BasTI}
        {Pietrinferni, A., Cassisi, S., Salaris, M., \& Castelli, F.} 2004, 
	 \textit{ApJ} 612, 168


\bibitem[Stanimirovic]{STANIMI99}
        {Stanimirovi\'c, S., Staveley-Smith, L., Dickey, J. M., Sault, R. J., \& Snowden, S. L.} 1999, 
	 \textit{MNRAS} 302, 417

 
	

\end{thebibliography}
\end{document}